\documentclass[%
 nofootinbib,reprint,
 amsmath,amssymb,
 aps,
prl,
superscriptaddress,
nofootinbib % <-- For compatibility with 
12pt]{revtex4-2}

\usepackage{comment}
\usepackage[caption=false]{subfig}
\usepackage{graphicx}% Include figure files
\usepackage{dcolumn}% Align table columns on decimal point
\usepackage{bm}% bold math
\usepackage{float}
\usepackage{tikz}
\usepackage{helvet}

\usepackage[show]{ed}
\newcommand{\ket}[1]{\left|#1\right\rangle}

\usepackage{siunitx}

\begin{document}

\preprint{APS/123-QED}

\title{Quantum Nonlocal Games on Graph Ensembles}

\author{Joshua Tucker}
\email{Contact Author: jt565@kent.ac.uk}
\author{Chris Weeks}
\affiliation{Quantum Applications Research Centre (QuARC), University of Kent, Ingram Building, Canterbury, Kent, CT2 7NH, United Kingdom}
\affiliation{Physics of Quantum \& Materials Research Group, School of Engineering, Mathematics \& Physics, University of Kent, Ingram Building, Canterbury, Kent, CT2 7NH, United Kingdom}
\author{Peter~Drmota}
\email{Contact Author: peter.drmota@physics.ox.ac.uk}
\author{Ellis~M.~Ainley}
\author{Ayush~Agrawal}
\author{Adam~R.~Martinez}
\author{Erin~Malinowski}
\author{Jacob~A.~Blackmore}
\author{David~P.~Nadlinger}
\email{Contact Author: david.nadlinger@physics.ox.ac.uk}
\author{Gabriel~Araneda}
\author{David~M.~Lucas}
\affiliation{Clarendon Laboratory, Department of Physics, University of Oxford, Oxford, OX1~3PU, United Kingdom}
\author{Carlos~Perez-Delgado}
\affiliation{Quantum Applications Research Centre (QuARC), University of Kent, Ingram Building, Canterbury, Kent, CT2 7NH, United Kingdom}
\affiliation{School of Computing, University of Kent,
Canterbury, Kent, CT2 7NF, United Kingdom}
\author{Paul Strange}
\author{Jorge Quintanilla}
\email{Contact Author: J.Quintanilla@kent.ac.uk}
\affiliation{Quantum Applications Research Centre (QuARC), University of Kent, Ingram Building, Canterbury, Kent, CT2 7NH, United Kingdom}
\affiliation{Physics of Quantum \& Materials Research Group, School of Engineering, Mathematics \& Physics, University of Kent, Ingram Building, Canterbury, Kent, CT2 7NH, United Kingdom}

\date{\today}% It is always \today, today,

%\begin{abstract}
%An article usually includes an abstract, a concise summary of the work
%covered at length in the main body of the article. 
%\end{abstract}

%\begin{abstract}
    
%\end{abstract}

\maketitle

%\ednote{-3750 words maximum }
%[Introduction (2 columns)]
%[DESCRIPTION - Introduction to the problems at hand] 
%2 Columns
%\ednote{Introduction (max. 2 full columns)}

{\bf
Quantum entanglement is one of the most striking discoveries in all of science. This effect allows, for instance, two spatially separated agents to coordinate their actions, without communication, to an extent that is both counter-intuitive, and provably impossible by any other physical means~\cite{cleve2004consequences}. A recently discovered example is that of mobile agents (players) performing spatial coordination tasks~\cite{brukner2006entanglement,Mironowicz_2023,PhysRevA.109.042201,Tucker2024quantum,WeeksPREPRINT} such as rendezvous, where the agents aim to meet on a network without communication. Until now, demonstrations of this advantage have relied on highly idealized conditions: agents are assumed to have complete knowledge of the topography, and experiments have been restricted to simulations using data generated by qubits within a single quantum processor. Here we address both limitations by developing a theory for graph ensembles that capture topographical uncertainty and by experimentally demonstrating the advantage in rendezvous scenarios between physically separated ion-trap systems with access to remote entanglement. Moreover, we simulate a broader set of problems on superconducting hardware. Surprisingly, when players are given the ability to gather more local information the quantum advantage increases -- a feat impossible by classical means. Our findings establish a concrete route toward practical quantum advantages in motion coordination problems. More broadly, they point to a new way of using portable quantum devices to enhance collective decision-making in uncertain environments.
}

Bipartite nonlocal quantum games involve two players, Alice and Bob, who share an entangled resource and interact with a referee, but are not allowed to communicate with each other during the game. The referee gives the players individual challenges with a joint goal and they may exploit their shared entanglement to coordinate their responses. The possibility of overcoming the limitations of the best classical strategy, i.e., realising operational quantum advantage, is closely related to violating Bell inequalities~\cite{Freedman1972,Aspect1982,Hensen2015,Giustina2015,Shalm2015}, which have many applications in quantum information processing (QIP), including quantum key distribution~\cite{Ekert1991,Arnon-Friedman2018,Vazirani2014} and certification of quantum processes~\cite{Colbeck2009,Pironio2010,Reichardt2013}.
In graph-based nonlocal games, the players' actions relate to assignments in an underlying graph, such as the colour of the vertices.
Experimental realisations of graph-based nonlocal games include the nonlocal Mermin–Peres magic-square game~\cite{Xu2022} and the odd-cycle game~\cite{Drmota2025}. In mobile-agent coordination games, the players' actions correspond to movements on the underlying graph.
It has been shown that players who share quantum entanglement can improve their chances of success by exploiting the correlated nature of local measurements performed on their part of the entangled state~\cite{brukner2006entanglement,Mironowicz_2023,PhysRevA.109.042201,Tucker2024quantum,WeeksPREPRINT}. Here, we demonstrate that this advantage extends to scenarios where the players face uncertainty with regard to the graph on which they will be placed. This addresses the reality of changing environments and uncertain circumstances that may apply to future applications.

\begin{figure*}
    \centering
    \includegraphics[width=1.0\textwidth]{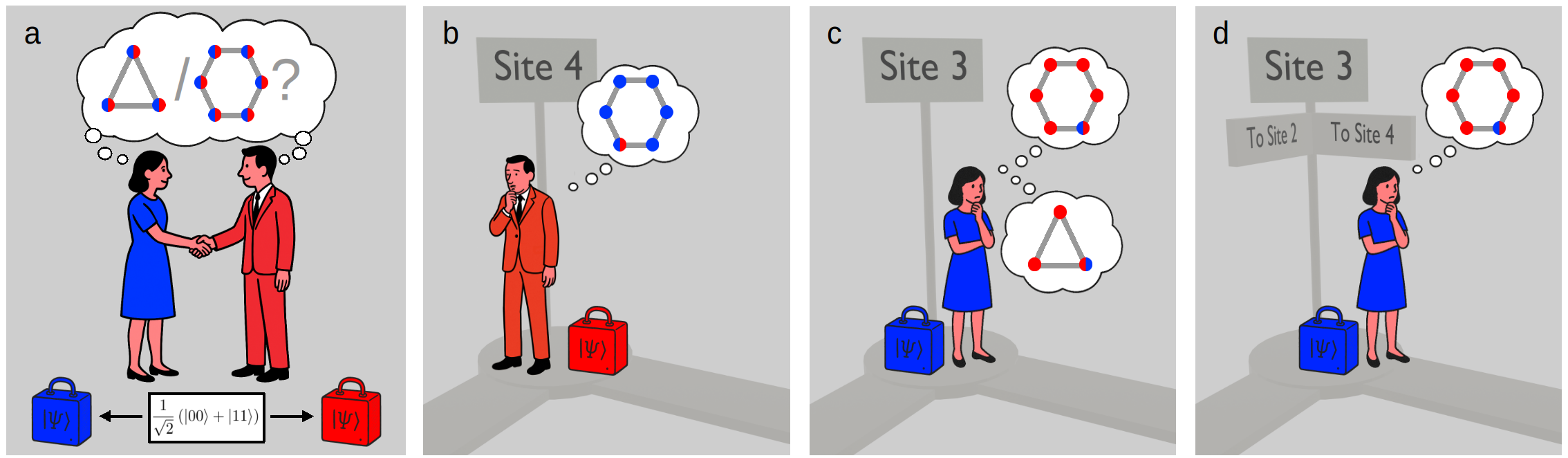} 
    \caption{\label{fig:game_illustration}%
        The games we consider involve two agents, Alice and Bob, coordinating their moves on a graph in order to achieve a common task. They are played in two phases: in phase one (a) Alice and Bob can agree a strategy and share quantum information without knowing their starting locations and with only statistical information about the graph's topology; in phase two (b-d) Alice and Bob have to make their moves without communicating and using only local information (see main text). 
    %One step rendezvous problem on the triangle graph $K_3$ (a) and the hexagon graph $C_6$ (b). In both examples, Alice and Bob start on sties 1 and 2, respectively, and move to the adjacent site with the highest index. However, this move leads to a successful rendezvous on the triangle but a failure on the hexagon. In this work we consider scenarios where the players do not know in advance which graph they will be placed on. This has to be taken into account by their joint strategy.%
    }
\end{figure*}
The mathematical models we consider here represent easy to understand real-world scenarios: two agents are introduced into an unknown territory, at unknown starting locations, without means of communication, and aim to meet each other (rendezvous~\cite{alpern2010rendezvous,Mironowicz_2023,Tucker2024quantum,PhysRevA.109.042201}), or cover the largest amount of territory (graph domination~\cite{PhysRevA.109.042201,WeeksPREPRINT}) using only local signposts and landmarks.
 
The game is played in two phases. In the first phase [Fig.~\ref{fig:game_illustration}(a)], Alice and Bob can communicate but they do not know their starting positions or the graph where the game will be played -- only the statistical ensemble from which the graph will be drawn. In our example, that ensemble consists of the labelled graphs $C_3$ and $C_6$, with equal probabilities. During this phase Alice and Bob may agree a strategy -- taking into account the statistical expectation of the topography -- and they may also establish entanglement between their quantum resources, if they have any. Then, the referee decides the graph and places the players on random vertices. In the second phase [Fig.~\ref{fig:game_illustration}~(b-d)], communication is no longer possible and each player acquires local information including the site they have been placed on. This information can reveal the topography in some cases. For example, if Bob is placed on site 4 [Fig.~\ref{fig:game_illustration}(b)], Bob learns that the game is played on $C_6$ because $C_3$ does not have a fourth site. In contrast, Alice may have landed on site 3 [Fig.~\ref{fig:game_illustration}(c)], which is common to both graphs. For cycle graphs, the players always have a two-way choice. We assume that Alice and Bob always know which direction leads to a vertex with higher index.\footnote{
    This is the formulation used in recent literature~\cite{Mironowicz_2023,PhysRevA.109.042201,Tucker2024quantum,WeeksPREPRINT} which dealt with single graphs. In that case, increasing or decreasing the cyclic index label provides an equivalent formulation. However, for statistical graph ensembles the two are \emph{not} equivalent. We stick to the higher/lower formulation as the language matches that used in earlier references. The other case is described in the Supplementary Information. None of our main conclusions depend on this choice.
} %
The players then move simultaneously and finally evaluate whether their objective was achieved.
If Alice and Bob share a pair of entangled qubits and are able to apply local unitary operations and measurements, they can decide their move on the basis of a measurement outcome (for instance, 1 = go-to-highest, 0 = go-to-lowest). The players may perform a different unitary operation depending on the information they can gather locally (including, but not limited to, the index $i$ of the vertex they find themselves in). This strategy introduces a new source of correlation between the players' moves that depends on the combination of sites they have been assigned. This correlation can lead to a higher probability of ending up on the same site, without any communication.  

In variants of such games, the availability of additional local information increases the players' ability to deduce the topography. For example, in Fig.~\ref{fig:game_illustration}d, a local signpost reveals the existence of a fourth site to Alice, who infers that the game is played on $C_6$. As we will see later, this variation can increase the collective quantum advantage, demonstrating that shared quantum resources allow the players to exploit local knowledge to further their joint goal without communicating.

Graph ensembles can be partially characterised by their entropy
\begin{equation}
    S = -\sum_{i=1}^N P(G_i) \log_2 P(G_i) \ ,
    \label{eq:ensemble_entropy}
\end{equation}
where $P(G_i)$ denotes the probability of a graph in the ensemble, $G_i$, being chosen.
$S=0$ when the players know the graph in advance (a single-graph ensemble~\cite{Mironowicz_2023,PhysRevA.109.042201,Tucker2024quantum}). Conversely, for a given $N$, $S$ is maximal when all $N$ graphs are equally likely. We will call $S$ the ``topographic uncertainty'', as it describes the lack of information Alice and Bob have as to the correct map of the terrain on which they will find themselves. This must be distinguished from other sources of uncertainty associated with the players' starting positions, any shared randomness, or the shared quantum state.

In general, the optimal strategies will change upon the introduction of topographic uncertainty, as winning moves for one graph in the ensemble may be losing moves for another, and vice versa. One key question is whether it is still possible to find quantum-assisted strategies that beat classical ones for such graph ensembles. As we will see, the answer is ``yes.''
Furthermore, when players land on the graph they may learn additional information (see Fig.~\ref{fig:game_illustration}).
However, by then, it is too late to change the joint strategy.
Classical strategies directly relate any available local information to a fixed move; quantum strategies, on the other hand, relate this information to a fixed basis of a quantum measurement whose outcome determines the move. This enables correlated moves that are provably impossible in a classical framework.

\section{Persistence of quantum advantage under topographic uncertainty}
Let us now turn our attention to the persistence of quantum advantage when $S\neq 0$. We will focus initially on the game of rendezvous~\cite{alpern2010rendezvous}. Specifically, we will consider a synchronous, one-step game with the following assumptions: the players may start on any site, they are not allowed to wait (they have to move), and they win if they end up on the same site or exchange places (meet on the way), but not simply by starting on the same site.\footnote{In the parlance of the recent literature~\cite{Mironowicz_2023,Tucker2024quantum} this is the $S=1,E=1,W=0,$ ``check-later''~\cite{Tucker2024quantum} version of the synchronous, 2-player, 1-step rendezvous game.} The game will be played in a statistical ensemble of $N$ graphs. We define $n$ as the number of sites in the largest graph of the ensemble.

The figure-of-merit for each strategy $\Sigma$ is defined as the expectation value of the winning probability,
\begin{equation}\label{eq:GeneralEnsembleEquation}
    P_W(\Sigma) = \sum_{i=1}^N P(G_i) P(W|\Sigma,G_i) \ ,
\end{equation}
which is the weighted sum of conditional probabilities $P(W|\Sigma,G_i)$ of winning a game on graph $G_i$ using the strategy $\Sigma$.
In the following, the symbol $\bar{P}_W$ will be used to distinguish experimental averages from the theoretical limit, $P_W$.

In the classical case, a joint strategy $\Sigma_c=\left(\sigma^A_1,\sigma^B_1,\sigma^A_2,\sigma^B_2,\ldots,\sigma^A_n,\sigma^B_n\right)$ prescribes a list of actions $\sigma^A_i$ and $\sigma^B_i$ to be taken by Alice and Bob respectively, when they land on site $i$. Each action can be ``go to the higher-indexed site'' ($\sigma_i = 1$) or ``go to the lower-indexed site'' ($\sigma_i = 0$). Allowing the instructions for each player to be different, the list contains $2n$ binary digits. As the optimal classical strategy is always deterministic, we do not consider probabilistic strategies here. Note that the strategy must be defined uniquely for a given ensemble, including any instructions for individual players to adapt their behaviour if they discover the graph they are on. However, such bifurcations can only be applied individually to each player on the basis of the information they have obtained locally; there can be no global bifurcation in the behaviour of both players on the basis of information available only to one of them, as that would require inter-player communication.

In the case of quantum strategies, Alice and Bob share a maximally entangled 2-qubit state (one qubit held by each player) such as the Bell state $\ket{\Phi^+} = \frac{1}{\sqrt{2}}\left(\ket{00}+\ket{11}\right).$ After landing on their starting site, each player applies a local unitary rotation to their qubit. They then measure their qubit in the computational basis and go to the higher-indexed (lower-indexed) site if the measurement outcome was 1 (0). A quantum strategy is therefore specified by $2n$ rotation angles, $\Sigma_q=\left(\theta^A_1,\theta^B_1,\theta^A_2,\theta^B_2,\ldots,\theta^A_n,\theta^B_n\right)$.

Previous studies of mobile-agent quantum-assisted coordination have considered only $S=0.$ To test whether quantum advantage survives the introduction of topographic uncertainty, we will consider first the simple ensemble described in Fig.~\ref{fig:game_illustration},
%case with only 2 graphs that are equally likely, leading to 
for which $S=1.$ 
%We choose the two cycle graphs shown in Fig.~\ref{fig:game_illustration}: $C_3$ and $C_6$. 
Both graphs in the ensemble, $C_3$ and $C_6$, have been considered individually in the literature~\cite{Mironowicz_2023,Tucker2024quantum}.
Finding optimal strategies involves optimizing the Boolean variables $\sigma_i$ in the classical case, and the rotation angles $\theta_i$ in the quantum case.
We find the optimal classical strategy $\Sigma_c^{\text{opt}}$ for rendezvous on ensembles involving \{$C_3$,$C_6$\} (see Methods), yielding a maximum classical rendezvous probability of $P_W(\Sigma_c^{\text{opt}}) \approx 0.5833$.
To search for near-optimal quantum strategies, we employ numerical optimisers. We present two particular strategies, $\Sigma_q^1$ and $\Sigma_q^2$, which are obtained by differential evolution~\cite{storn1997differential} and a Limited-memory Broyden–Fletcher–Goldfarb–Shanno algorithm (L-BFGS-B~\cite{doi:10.1137/0916069}), with rendezvous probabilities of $P_W(\Sigma_q^1) \approx 0.6043$ and $P_W(\Sigma_q^2) \approx 0.6086$, respectively (see Methods).
Thus, both quantum strategies outperform the optimal classical strategy, demonstrating the persistence of quantum advantage in the presence of topographic uncertainty. 

\begin{figure}[h]
    \includegraphics[]{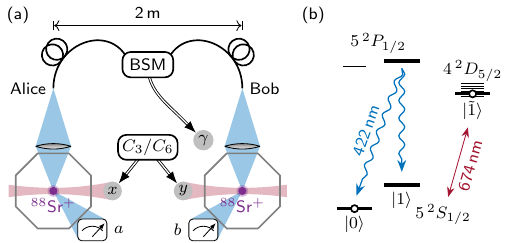}
    \caption{(a) The experimental apparatus used to implement rendezvous games on graph ensembles consists of two trapped ${}^{88}\mathrm{Sr}^{+}$ ions. The ions can be entangled remotely by a Bell state measurement (BSM) on two photons that are each entangled with one of the ions.
    The BSM heralds the creation of the state in Eq.~\eqref{eq:raw-entanglement} whose phase  $\vartheta_\gamma$ is communicated to Bob.
    The variables $x$ and $y$ encode the sites at which Alice and Bob have been placed on the graph and the players perform unitary rotations and projective measurements using lasers, yielding binary outcomes $a$ and $b$, respectively.
    (b) Relevant electronic level structure of ${}^{88}\mathrm{Sr}^{+}$.
    }\label{fig:oxford}
\end{figure}
To test our predictions experimentally, we use two ${}^{88}\mathrm{Sr}^{+}$ ions, each trapped in physically separate ion trap systems (Alice and Bob) approximately \SI{2}{\meter} apart~\cite{stephenson_high-rate_2020}, and controlled independently (see Fig.~\ref{fig:oxford}a). The qubit $\{\ket{0}, \ket{1}\}$ is encoded in the Zeeman sub-levels of the ground state.
To generate remote entanglement, a synchronised laser pulse excites each ion to the short-lived $P_{1/2}$ state, from which it decays spontaneously into the qubit subspace via emission of a 422-nm photon whose polarisation is entangled with the final qubit state. These photons are routed through optical fibres to a central Bell state analyzer, which signals when one of four two-photon coincidence patterns, $\gamma \in \{0,1,2,3\}$, is detected. This detection heralds the creation of remote entanglement between the qubits of the form
\begin{equation}
    \ket{\Psi^{+}(\vartheta_\gamma)} = \frac1{\sqrt{2}}\left(\ket{01}+\mathrm{e}^{\mathrm{i}\vartheta_\gamma}\ket{10}\right) \ , \label{eq:raw-entanglement}
\end{equation}
where the angles $\vartheta_\gamma$ are measured independently and are compensated in the experiment by application of a phase gate to Bob's qubit. After the control systems gain knowledge of the site on which they were placed in the unknown graph, the qubits are manipulated in real-time using a narrow-linewidth laser, capable of driving the dipole-forbidden $S_{1/2}\leftrightarrow D_{5/2}$ transition, to implement the unitary prescribed by the quantum strategy $\Sigma_q^2$. Local state-dependent fluorescence measurements determine the action of the players and the final locations of Alice and Bob are exchanged to evaluate the rendezvous probability, yielding $\bar{P}_W(\Sigma_q^2) = 0.5992(8)$.
Thus, the experimental result exceeds the classical limit by 20 standard deviations, which we interpret as a distinct advantage to the quantum strategy.
The difference between the observed winning probability and the theoretical maximum can be explained by errors in the remotely entangled state, which we characterise using quantum state tomography. The maximum likelihood estimate of the fidelity to the closest maximally entangled state is $0.971(3)$, and the fidelity to $\ket{\Phi^{+}}$ is  $0.965(3)$, indicating that applying coherent rotations could improve the winning probability. The remaining infidelity is dominated by imperfections in the single-photon collection; the ion state is also affected by magnetic field noise, laser phase noise, and residual entanglement with the motion. 
Using the maximum likelihood estimate of the experimental density matrix, we numerically simulate the rendezvous protocol and obtain $\bar{P}_W(\Sigma_q^2) = 0.5985(9)$, which is consistent with the observed winning probability.

Furthermore, quantum simulations of $\Sigma_q^1$ using a NISQ processor, IBM Kingston~\cite{IBMQuantum_Kingston} (see Methods) yield a rendezvous probability of $\bar{P}_W(\Sigma_q^1) = 0.5939(5)$, which also exceeds the optimal classical strategy by $21\sigma$.
This result falls short of the theoretical prediction due to hardware imperfections, which were not further characterised.

\section{Increased quantum advantage with more local information}

To further explore the mechanism of quantum advantage under topographic uncertainty, we modify the model to give players access to additional local information that they can use to deduce the graph they are on. Specifically, we allow players to know the labels of the two sites that are accessible to them  (Fig.~\ref{fig:game_illustration}d). In our earlier example, players knew they were in graph $C_6$ if their allocated site was 4,5, or 6. In contrast, when placed on sites 1, 2, or 3 they could not deduce whether they were on $C_3$ or $C_6$. Since sites 4-6 are not present in $C_3$, this information was only used implicitly in their strategies (whether classical or quantum). With the additional signposts, players can deduce the graph also when they are placed on site 1 (which is connected to 2 and 3 on $C_3$, and to 2 and 6 on $C_6$) or 3 (connected to 1 and 2 on $C_3$, and to 2 and 4 on $C_6$). This can be taken into account explicitly in a rendezvous strategy. In practical terms, this means introducing two additional variables per player in the specification of the strategy (Boolean variables in the classical case or rotation angles for a quantum strategy). For instance, the single variable specifying the action of a player after landing on site 1 is replaced with one variable for the case when the player has deduced they are on $C_3$ and another for $C_6$. Likewise, the variable describing what the player does when landing on site 3 also becomes two distinct variables. All variables then need to be optimized -- see Methods. 

We simulate rendezvous probabilities in the presence of signposts using an optimal classical strategy and a quantum strategy, $\Sigma_q^3$, optimized using L-BFGS-B~\cite{doi:10.1137/0916069}, which yields $P_W(\Sigma_q^3) = 0.6250$.
This value is the average of the individual optimal strategies for each graph in  the ensemble~\cite{Mironowicz_2023}. Thus, the additional information enables a general strategy that maximizes the probability of winning on each graph.
The classical prediction is the same as in the previous model because $\Sigma_c^{\text{opt}}$ is optimal for the individual graphs in the ensemble, so the players get no additional benefit in deviating from the original strategy.
Defining the quantum advantage in rendezvous by
\begin{equation}
    Q=P_W(\Sigma_q)-P_W(\Sigma_c^\mathrm{opt}),
    \label{eq:quantum_advantage_rendezvous}
\end{equation}  
this result corresponds to a ${Q_\text{See Labels}}/{Q_\text{Can't See Labels}} - 1 = \SI{65}{\percent}$ increase in the quantum advantage, compared to the best quantum strategy we found for the case when the additional local information was not available, $\Sigma_q^2$. 

In simulations using IBM Kingston, we obtain $\bar{P}_W(\Sigma_q^3) = 0.6147(5)$ %which exceeds the optimal classical strategy with $66\sigma$ confidence.
which exceeds the bound set by the optimal classical strategy by $66\sigma$.
The NISQ simulation also shows considerably better performance when more local information is available -- indeed, it is $\SI{24}{\percent}$ above the theoretical prediction for $\Sigma_q^2$.

\begin{figure*}[!ht]
    \centering
    %
    % First Subfigure
    \subfloat{%
        \includegraphics[width=0.48\linewidth]{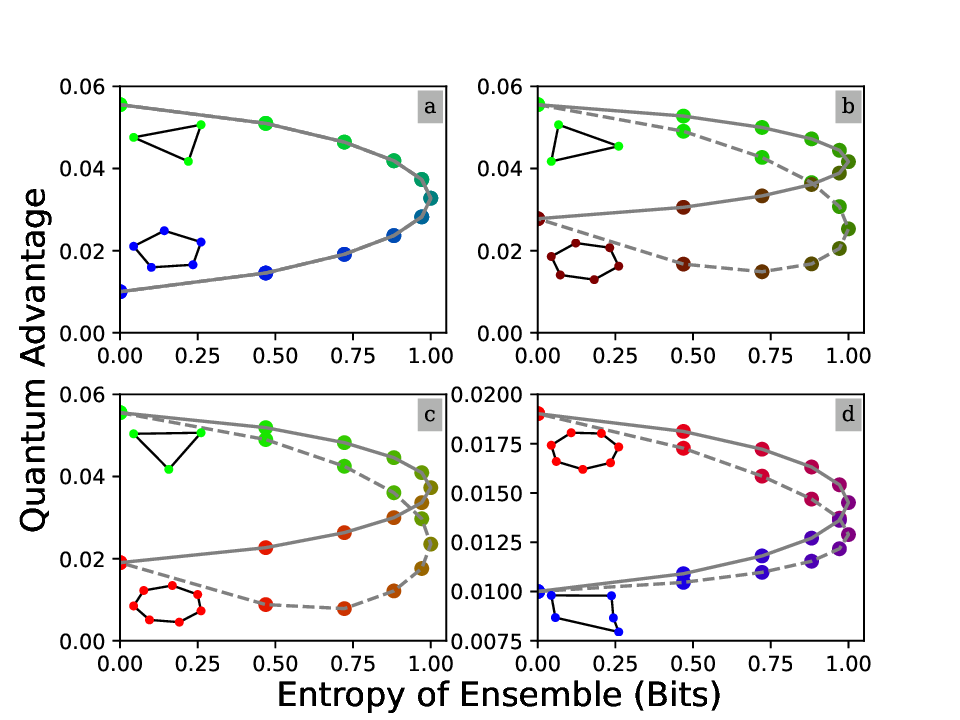}%
    }\hfill
    % Second Subfigure
    \subfloat{%
        \includegraphics[width=0.48\linewidth]{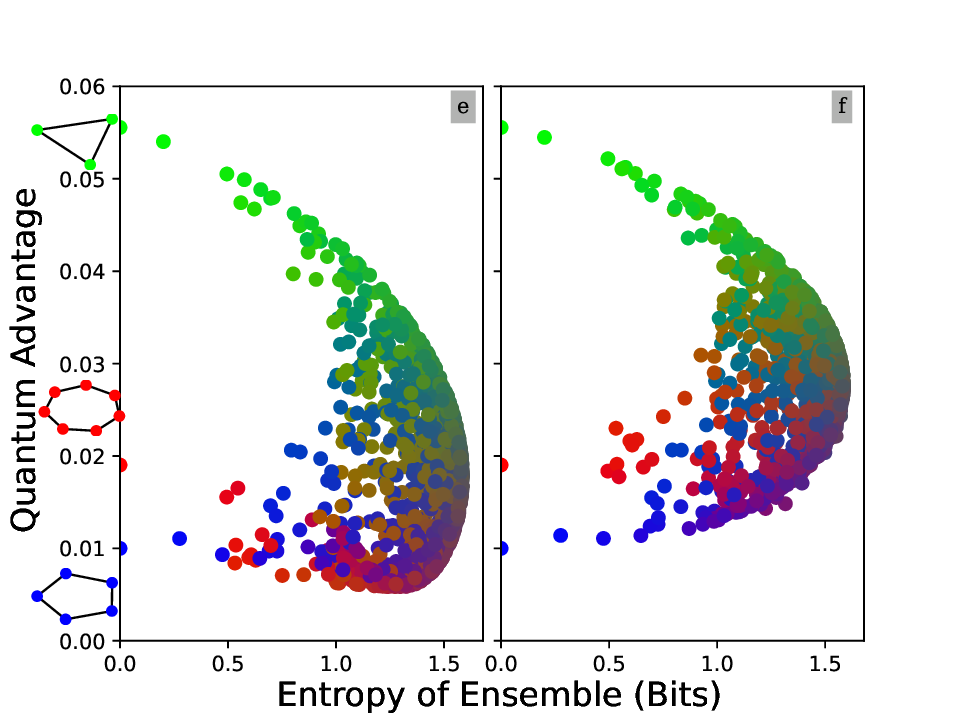}%
    }
    \caption{Quantum advantage $Q$ [Eq.~\eqref{eq:quantum_advantage_rendezvous}] -- defined as the increase in rendezvous probability when using a quantum strategy over the best classical strategy -- for two-player, one-step rendezvous in an uncertain topography, represented by a statistical ensemble of the graphs  \{$C_3$,$C_5$\} (a), \{$C_3$,$C_6$\} (b), \{$C_3$,$C_7$\} (c), \{$C_5$,$C_7$\} (d), and \{$C_3$,$C_5$,$C_7$\} (e-f). The horizontal axis shows the entropy $S$ of each  ensemble [Eq.~(\ref{eq:ensemble_entropy})]. This quantifies the players' initial topographic uncertainty, before they are placed on their starting site. For each ensemble, the two scenarios described in Fig.~\ref{fig:game_illustration} are considered: one where less local information is available to the players, shown by the circles joined by dashed lines in panels a-d and by the circles in panel e; and a variant with more local information available, shown by the circles joined by solid lines in a-d and the circles in f. Each graph is assigned a colour as indicated on the panels; to visualise the composition of each ensemble, these colours are mixed accordingly. The curves in panels a-d give at most two, discrete values of $Q$ for each value of the entropy, corresponding to the pair of 2-graph ensembles that realise that amount of entropy. In contrast, panels e and f show a continuum of $Q$ values. These correspond to the continuum of different 3-graph ensembles that have the same entropy.}
    \label{fig:EntropyAgainstAdvantageTwoGraphEnsembles}
    \label{fig:EntropyAgainstThreeGraphEnsembles}
    \label{fig:EntropyAgainstGraphEnsembles}
\end{figure*}

To establish the generality of this remarkable increase in quantum advantage upon giving players access to additional local information we extend our calculations to other families of graph ensembles: \{$C_3$,$C_5$\}, \{$C_3$,$C_6$\}, \{$C_3$,$C_7$\}, \{$C_5$,$C_7$\},  and \{$C_3$,$C_5$,$C_7$\} (for each set of graphs, a family of ensembles is obtained by varying the probability distribution of the graphs in the set). The results are shown in Fig.~\ref{fig:EntropyAgainstGraphEnsembles}. 
 In all ensembles except \{$C_3$,$C_5$\}, the quantum advantage is greater in the case with more local information, indicating that players can exploit this additional knowledge to improve their winning chances. Crucially, this appears to be a purely quantum effect: in the classical case, winning probabilities remain unchanged regardless of whether players have access to this additional local information. This is true even in ensembles including the graph $C_7$ for which the single-graph, optimal classical strategy differs from those of the other graphs (this is generic -- the Supplementary Information contains additional data including more graphs with different classical strategies). The exception for \{$C_3$,$C_5$\} occurs because players can already employ a joint strategy that achieves the theoretical maximum winning probability for each graph individually, leaving no room for further advantage. 
 
 The ability of quantum strategies to make better use of additional local information suggests, counter-intuitively, an additional source of quantum advantage in uncertain environments. 

\section{Extension to graph domination games}
We now consider the graph domination game~\cite{PhysRevA.109.042201,WeeksPREPRINT}. It is played identically to the rendezvous game except that, instead of attempting to meet, the players aim to maximize the number $D$ of unique vertices ``dominated,'' that is, sites that contain a player or are neighbours to one of these via an edge. The figure of merit in this case is  `domination number' (average value of $D$), which we  normalise by the number of sites averaged over the graphs in the ensemble, $\bar{N}$. Comparing this quantity for optimised quantum and classical strategies yields a measure of quantum advantage analogous to (\ref{eq:quantum_advantage_rendezvous}): 
\begin{equation}
    Q=
    \frac{1}{\bar{N}}
    \left(D^{\text{quantum}}-D^{\text{classical}}\right).
    \label{eq:quantum_advantage_domination}
\end{equation} 
Optimal strategies for the graph domination game are usually player-asymmetric, due to the large advantage in keeping the players' movements opposite when they start on the same site. 

\begin{figure}[ht]
    \centering
    \includegraphics[width=0.4\textwidth]{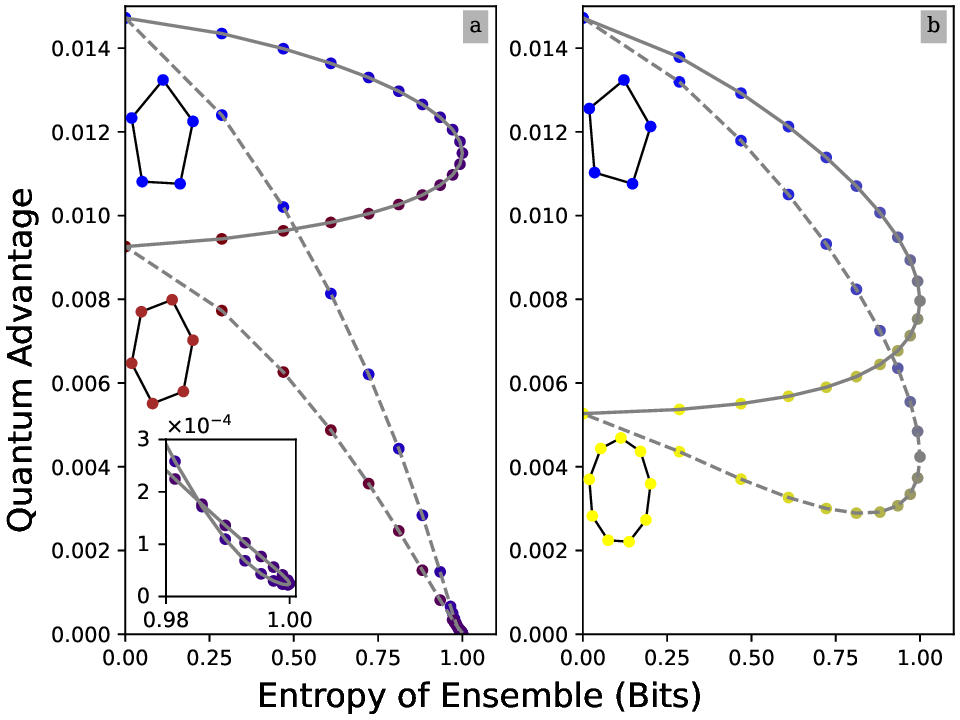}\\[1em] % <-- extra space
    \caption{Quantum advantage for one-step graph domination game played on the \{$C_5$,$C_6$\} (a) and \{$C_5$,$C_9$\} (b) families of graph ensembles. The circles joined by dashed lines correspond to the case with less local information (Fig.~\ref{fig:game_illustration}c) and the circles joined by solid lines show the case with more local information (Fig.~\ref{fig:game_illustration}d). The positive quantum advantage and its increase when more local information is available mirror our results for rendezvous games.
    }
    \label{fig:MultiEnsembleDomination}
\end{figure}

We investigate two families of ensembles:\footnote{$C_5$ is the smallest cycle on which the domination game is non-trivial.} \{$C_5$,$C_6$\} and \{$C_5$,$C_9$\}, again exploring ensembles with different topographic entropies $S$ and comparing the quantum advantages with the same two different types of local information considered for rendezvous. The results, shown in Fig.~\ref{fig:MultiEnsembleDomination}, exhibit features analogous to the rendezvous games described above: i) the quantum advantage is always positive, in spite of the topographic uncertainty; ii) the quantum advantage increases when players have access to greater local information. The \{$C_5$,$C_6$\} ensembles with high topographic entropy present a particularly stark case: with less local information, their quantum advantage is almost zero, meaning that the best quantum strategy we could find is only slightly better than the optimal classical strategy, and the quantum advantage is consequently much lower than for the two corresponding $S=0$ single-graph results; in contrast, with more local information the quantum advantage jumps to values above that of the $C_6$, $S=0$ ensemble. We note that in this case the optimal classical strategy for the ensemble is also the optimal classical strategy for the individual graphs, leaving no room for the classical strategy to improve upon the introduction of additional local information (similar to what happened for rendezvous in \{$C_3$,$C_6$\}). 

It can be seen that for \{$C_5$,$C_6$\}, when the ensemble has the highest uncertainty and the players are unable to see the nodes that connect to their current location, the quantum advantage reaches a minimum. This is a result we do not see within rendezvous. Also unique to domination, we find that the lines cross, resulting in two unique ensembles that have the same quantum advantage. For ensembles with {$C_5$,$C_9$} we find performance similar to that of rendezvous games on ensembles.%, showing the close link between these games.

In summary, our numerical analysis of graph-domination mobile-agent games supports all the main conclusions we reached from our study of rendezvous: quantum advantage survives topographic uncertainty and it increases when players have access to additional local information. 

\section{Discussion}

We have shown that there is a systematic increase of quantum advantage when players are given access to additional local information (see Table~\ref{tab:max_rel_increase} for a summary of maximum enhancement for each game and family of ensembles). Suppose that Alice lands on a site where some such additional information is available, while Bob lands on a site where this is not the case. In the classical case, Alice can use the additional information to inform their actions, but this has no bearing on Bob's actions. In the quantum case, Alice uses the extra information to change the measurement basis, which influences the correlations shared with Bob. So, although Alice cannot \emph{transmit} this information to Bob, Alice and Bob can exploit it to improve the coordination of their moves. Crucially, this is a purely quantum effect, as classical strategies cannot benefit from additional local information gained after the players have been separated. We suggest this as a useful conceptualization of the origin of quantum advantage in a wide variety of nonlocal games.
%
\begin{comment}
    \begin{table}
        \centering
        \begin{tabular}{c|c}
             Game and Ensemble & Max($\frac{\Delta F^n}{F^n_c} - \frac{\Delta F^{nn}}{F^{nn}_c} $)\\
             \hline
            Rendezvous \{$C_3$,$C_5$\}& 0\\
            Rendezvous \{$C_3$,$C_6$\} & 0.0396 \\
            Rendezvous  \{$C_3$,$C_7$\} &0.0427 \\
            Rendezvous \{$C_5$,$C_7$\} & 0.0044\\
            Rendezvous  \{$C_3$,$C_5,C_7$\}& 0.0444 \\
            Graph Domination \{$C_5$,$C_6$\}& 0.0134\\
            Graph Domination \{$C_5$,$C_9$\}& 0.0056\\
        \end{tabular}
        \caption{Caption}
        \label{tab:placeholder}
    \end{table}
\end{comment}
%
\begin{table}
    \centering
    \begin{tabular}{c|c|c}
         Game & Ensembles & $\max{\Delta Q/Q^0}$ \\
         \hline
        Rendezvous & \{$C_3$,$C_5$\}& 0\\
        Rendezvous &\{$C_3$,$C_6$\} & 1.242 \\
        Rendezvous & \{$C_3$,$C_7$\} & 2.360 \\
        Rendezvous& \{$C_5$,$C_7$\} & 0.126 \\
        Rendezvous & \{$C_3$,$C_5,C_7$\}& 3.035 \\
        Domination& \{$C_5$,$C_6$\}& 459.129 \\
        Domination& \{$C_5$,$C_9$\}& 1.208 \\
    \end{tabular}
    \caption{Maximum relative increase of quantum advantage when players are allowed to access additional local information. The additional local information are the  signposts shown in Fig.~\ref{fig:game_illustration}d. For each family of graph ensembles, different probability distributions are examined and for each of those the quantum advantage is computed with signposts ($Q^1$) and without signposts ($Q^0$). The source data are those in Fig.~\ref{fig:EntropyAgainstAdvantageTwoGraphEnsembles} and Fig.~\ref{fig:MultiEnsembleDomination}. We calculate the relative increase $\left(Q^1-Q^0\right)/Q^0$ and find the ensemble, within each family, for which this quantity is maximum. The large value for the \{$C_5$,$C_6$\} ensemble is due to the advantage in the case of less local information being close to zero at maximum ensemble entropy.}
    \label{tab:max_rel_increase}
\end{table}

Our results imply that academic discussions of graph-based nonlocal games (where typically players know perfectly the graph the game is based on, and have limited ability to obtain additional information as the game progresses) may underestimate the benefit of quantum strategies considerably. In real-world situations, topography or, more generally, environmental circumstances, are changing and uncertain. We conjecture that quantum strategies may allow us to compensate for this by using sensors to obtain local information and use it in conjunction with shared quantum resources to maintain coordination.

Our results point to a fertile new field of application of quantum technologies in Operational Research~\cite{OverviewOR}. Graph ensembles describe many real systems --  for instance, small-world networks are statistical graph ensembles of mixed degree~\cite{Watts1998}.

Mobile-agent quantum games have been simulated before using NISQ processors~\cite{Tucker2024quantum,WeeksPREPRINT} but they have never been tested experimentally. Our experiments using trapped ions are thus the first experimental realisation of mobile-agent games generally, not just graph ensembles. On the other hand, the simulations using NISQ processors open the door to future quantum simulations of complex games involving large numbers of players. 

\section{Acknowledgements}
JT and CW acknowledge a studentships awarded by the Engineering and Physical Sciences Research Council (EPSRC) EP/W52461X/1 (Grant Numbers: 2872645 and 2923510). JT wishes to thank the Joseph Chapman, Lee Armstrong and Diksha Gautam for helpful discussions regarding the theory. JQ wishes to thank Elham Kashefi for a brief, but singularly helpful discussion. This work was supported by the National Quantum Computing Centre through its Quantum Computing Access Program (project number ACA0029). We acknowledge the use of IBM Quantum services for this work.
This work was supported by the U.K. EPSRC ``QCI3'' Hub EP/Z53318X/1. We thank Sandia National Laboratories for supplying the ion traps used in the experiment, and the developers of the experimental control system ARTIQ~\cite{artiq}. JAB acknowledges support from the EPSRC quantum technology career acceleration fellowship UKRI1223. PD is a director of and partially employed by Quantum Fabrix Ltd. The views expressed are those of the authors, and do not reflect the official policy or position of IBM or the IBM Quantum team.
\bibliography{apssamp}% Produces the bibliography via BibTeX.

@article{PhysRevA.109.042201,
  title = {Quantum strategies for rendezvous and domination tasks on graphs with mobile agents},
  author = {Viola, Giuseppe and Mironowicz, Piotr},
  journal = {Phys. Rev. A},
  volume = {109},
  issue = {4},
  pages = {042201},
  numpages = {15},
  year = {2024},
  month = {Apr},
  publisher = {American Physical Society},
  doi = {10.1103/PhysRevA.109.042201},
  url = {https://link.aps.org/doi/10.1103/PhysRevA.109.042201}
}

@article{Mironowicz_2023,
doi = {10.1088/1367-2630/acb22d},
url = {https://doi.org/10.1088/1367-2630/acb22d},
year = {2023},
month = {jan},
publisher = {IOP Publishing},
volume = {25},
number = {1},
pages = {013023},
author = {Mironowicz, P},
title = {Entangled rendezvous: a possible application of {{Bell}} non-locality for mobile agents on networks},
journal = {New Journal of Physics}
}

@article{Drmota2025,
  title = {Experimental Quantum Advantage in the Odd-Cycle Game},
  author = {Drmota, P. and Main, D. and Ainley, E. M. and Agrawal, A. and Araneda, G. and Nadlinger, D. P. and Nichol, B. C. and Srinivas, R. and Cabello, A. and Lucas, D. M.},
  journal = {Phys. Rev. Lett.},
  volume = {134},
  issue = {7},
  pages = {070201},
  numpages = {6},
  year = {2025},
  month = {Feb},
  publisher = {American Physical Society},
  doi = {10.1103/PhysRevLett.134.070201},
  url = {https://link.aps.org/doi/10.1103/PhysRevLett.134.070201}
}

@article{stephenson_high-rate_2020,
    title = {{High-Rate, High-Fidelity Entanglement of Qubits Across an Elementary Quantum Network}},
    author = {Stephenson, L. J. and Nadlinger, D. P. and Nichol, B. C. and An, S. and Drmota, P. and Ballance, T. G. and Thirumalai, K. and Goodwin, J. F. and Lucas, D. M. and Ballance, C. J.},
    journal = {Phys. Rev. Lett.},
    volume = {124},
    issue = {11},
    pages = {110501},
    numpages = {6},
    year = {2020},
    month = {Mar},
    publisher = {American Physical Society},
    doi = {10.1103/PhysRevLett.124.110501},
    url = {https://link.aps.org/doi/10.1103/PhysRevLett.124.110501}
}

@misc{ARTIQ,
    author = {Bourdeauducq, S{\'{e}}bastien and others},
    doi = {10.5281/zenodo.1492176},
    title = {{m-labs/artiq: 6.0 (Version 6.0)}},
    year = {2021}
}

@article{brukner2006entanglement,
  title={Entanglement-assisted orientation in space},
  author={Brukner, {\v{C}}aslav and Paunkovi{\'c}, Nikola and Rudolph, Terry and Vedral, Vlatko},
  journal={International Journal of Quantum Information},
  volume={4},
  number={02},
  pages={365--370},
  year={2006},
  publisher={World Scientific}
}

@article{WeeksPREPRINT,
  title = {Quantum-assisted domination games on cycle graphs},
  author = {Weeks, C. and Strange, P. and Drmota, P. and Quintanilla, J.},
  journal = {New Journal of Physics},
  year = {2026},
  doi = {10.1088/1367-2630/ae73a9},
  note = {In press}
}

@article{alpern2010rendezvous,
  title={Rendezvous search games},
  author={Alpern, Steve},
  journal={Wiley Encyclopedia of Operations Research and Management Science},
  year={2010},
  publisher={Wiley Online Library}
}

@inproceedings{cleve2004consequences,
  title={Consequences and limits of nonlocal strategies},
  booktitle={Proceedings. 19th IEEE Annual Conference on Computational Complexity, 2004.},
  pages={236--249},
  author = {Cleve, Richard AND Hoyer, Peter AND Toner, Ben AND Watrous, John},
  year={2004},
  organization={IEEE}
}

@article{OverviewOR,
author = {Fotios Petropoulos and Gilbert Laporte and Emel Aktas and Sibel A. Alumur and Claudia Archetti and Hayriye Ayhan and Maria Battarra et al.},
title = {Operational Research: methods and applications},
journal = {Journal of the Operational Research Society},
volume = {75},
number = {3},
pages = {423--617},
year = {2024},
publisher = {Taylor \& Francis},
doi = {10.1080/01605682.2023.2253852},
URL = { 
        https://doi.org/10.1080/01605682.2023.2253852
},
eprint = { 
        https://doi.org/10.1080/01605682.2023.2253852
}
}

@article{storn1997differential,
  title={Differential evolution--a simple and efficient heuristic for global optimization over continuous spaces},
  author={Storn, Rainer and Price, Kenneth},
  journal={Journal of global optimization},
  volume={11},
  number={4},
  pages={341--359},
  year={1997},
  publisher={Springer}
}

@article{doi:10.1137/0916069,
author = {Byrd, Richard H. and Lu, Peihuang and Nocedal, Jorge and Zhu, Ciyou},
title = {A Limited Memory Algorithm for Bound Constrained Optimization},
journal = {SIAM Journal on Scientific Computing},
volume = {16},
number = {5},
pages = {1190-1208},
year = {1995},
doi = {10.1137/0916069},

URL = { 
    
        https://doi.org/10.1137/0916069   

},
eprint = { 
    
        https://doi.org/10.1137/0916069
    
    

}

}

@article{Freedman1972,
  title = {Experimental Test of Local Hidden-Variable Theories},
  author = {Freedman, Stuart J. and Clauser, John F.},
  journal = {Phys. Rev. Lett.},
  volume = {28},
  issue = {14},
  pages = {938--941},
  numpages = {0},
  year = {1972},
  month = {Apr},
  publisher = {American Physical Society},
  doi = {10.1103/PhysRevLett.28.938},
  url = {https://link.aps.org/doi/10.1103/PhysRevLett.28.938}
}

@article{Aspect1982,
  title = {Experimental Test of {Bell}'s Inequalities Using Time-Varying Analyzers},
  author = {Aspect, Alain and Dalibard, Jean and Roger, G\'erard},
  journal = {Phys. Rev. Lett.},
  volume = {49},
  issue = {25},
  pages = {1804--1807},
  numpages = {0},
  year = {1982},
  month = {Dec},
  publisher = {American Physical Society},
  doi = {10.1103/PhysRevLett.49.1804},
  url = {https://link.aps.org/doi/10.1103/PhysRevLett.49.1804}
}

@Article{Hensen2015,
  author   = {Hensen, B. and Bernien, H. and Dréau, A. E. and Reiserer, A. and Kalb, N. and Blok, M. S. and Ruitenberg, J. and Vermeulen, R. F. L. and Schouten, R. N. and Abellán, C. and Amaya, W. and Pruneri, V. and Mitchell, M. W. and Markham, M. and Twitchen, D. J. and Elkouss, D. and Wehner, S. and Taminiau, T. H. and Hanson, R.},
  title    = {Loophole-free {Bell} inequality violation using electron spins separated by 1.3 kilometres},
  doi      = {10.1038/nature15759},
  issn     = {1476-4687},
  number   = {7575},
  pages    = {682--686},
  url      = {https://doi.org/10.1038/nature15759},
  volume   = {526},
  journal  = {Nature},
  refid    = {Hensen2015},
  year     = {2015},
}

@article{Giustina2015,
  title = {Significant-Loophole-Free Test of {Bell}'s Theorem with Entangled Photons},
  author = {Giustina, Marissa and Versteegh, Marijn A. M. and Wengerowsky, S\"oren and Handsteiner, Johannes and Hochrainer, Armin and Phelan, Kevin and Steinlechner, Fabian and Kofler, Johannes and Larsson, Jan-\AA{}ke and Abell\'an, Carlos and Amaya, Waldimar and Pruneri, Valerio and Mitchell, Morgan W. and Beyer, J\"orn and Gerrits, Thomas and Lita, Adriana E. and Shalm, Lynden K. and Nam, Sae Woo and Scheidl, Thomas and Ursin, Rupert and Wittmann, Bernhard and Zeilinger, Anton},
  journal = {Phys. Rev. Lett.},
  volume = {115},
  issue = {25},
  pages = {250401},
  numpages = {7},
  year = {2015},
  month = {Dec},
  publisher = {American Physical Society},
  doi = {10.1103/PhysRevLett.115.250401},
  url = {https://link.aps.org/doi/10.1103/PhysRevLett.115.250401}
}

@article{Shalm2015,
  title = {Strong Loophole-Free Test of Local Realism},
  author = {Shalm, Lynden K. and Meyer-Scott, Evan and Christensen, Bradley G. and Bierhorst, Peter and Wayne, Michael A. and Stevens, Martin J. and Gerrits, Thomas and Glancy, Scott and Hamel, Deny R. and Allman, Michael S. and Coakley, Kevin J. and Dyer, Shellee D. and Hodge, Carson and Lita, Adriana E. and Verma, Varun B. and Lambrocco, Camilla and Tortorici, Edward and Migdall, Alan L. and Zhang, Yanbao and Kumor, Daniel R. and Farr, William H. and Marsili, Francesco and Shaw, Matthew D. and Stern, Jeffrey A. and Abell\'an, Carlos and Amaya, Waldimar and Pruneri, Valerio and Jennewein, Thomas and Mitchell, Morgan W. and Kwiat, Paul G. and Bienfang, Joshua C. and Mirin, Richard P. and Knill, Emanuel and Nam, Sae Woo},
  journal = {Phys. Rev. Lett.},
  volume = {115},
  issue = {25},
  pages = {250402},
  numpages = {10},
  year = {2015},
  month = {Dec},
  publisher = {American Physical Society},
  doi = {10.1103/PhysRevLett.115.250402},
  url = {https://link.aps.org/doi/10.1103/PhysRevLett.115.250402}
}

@article{Xu2022,
  title = {Experimental Demonstration of Quantum Pseudotelepathy},
  author = {Xu, Jia-Min and Zhen, Yi-Zheng and Yang, Yu-Xiang and Cheng, Zi-Mo and Ren, Zhi-Cheng and Chen, Kai and Wang, Xi-Lin and Wang, Hui-Tian},
  journal = {Phys. Rev. Lett.},
  volume = {129},
  issue = {5},
  pages = {050402},
  numpages = {6},
  year = {2022},
  month = {Jul},
  publisher = {American Physical Society},
  doi = {10.1103/PhysRevLett.129.050402},
  url = {https://link.aps.org/doi/10.1103/PhysRevLett.129.050402}
}

@article{Ekert1991,
  title = {Quantum cryptography based on {Bell}'s theorem},
  author = {Ekert, Artur K.},
  journal = {Phys. Rev. Lett.},
  volume = {67},
  issue = {6},
  pages = {661--663},
  numpages = {0},
  year = {1991},
  month = {Aug},
  publisher = {American Physical Society},
  doi = {10.1103/PhysRevLett.67.661},
  url = {https://link.aps.org/doi/10.1103/PhysRevLett.67.661}
}

@article{Vazirani2014,
	author = {Vazirani, Umesh and Vidick, Thomas},
	title = {{Fully Device-Independent Quantum Key Distribution}},
	journal = {Phys. Rev. Lett.},
	volume = {113},
	number = {14},
	pages = {140501},
	year = {2014},
	month = sep,
	publisher = {American Physical Society},
	doi = {10.1103/PhysRevLett.113.140501}
}

@article{Arnon-Friedman2018,
	author = {Arnon-Friedman, Rotem and Dupuis, Fr{\ifmmode\acute{e}\else\'{e}\fi}d{\ifmmode\acute{e}\else\'{e}\fi}ric and Fawzi, Omar and Renner, Renato and Vidick, Thomas},
	title = {{Practical device-independent quantum cryptography via entropy accumulation}},
	journal = {Nat. Commun.},
	volume = {9},
	number = {459},
	pages = {459},
	year = {2018},
	month = jan,
	issn = {2041-1723},
	publisher = {Nature Publishing Group},
	doi = {10.1038/s41467-017-02307-4}
}

@article{Colbeck2009,
	author = {Colbeck, Roger},
	title = {{Quantum And Relativistic Protocols For Secure Multi-Party Computation}},
	journal = {arXiv},
	year = {2009},
	month = nov,
	eprint = {0911.3814},
	doi = {10.48550/arXiv.0911.3814}
}

@article{Pironio2010,
	author = {Pironio, S. and Ac{\ifmmode\acute{\imath}\else\'{\i}\fi}n, A. and Massar, S. and de la Giroday, A. Boyer and Matsukevich, D. N. and Maunz, P. and Olmschenk, S. and Hayes, D. and Luo, L. and Manning, T. A. and Monroe, C.},
	title = {{Random numbers certified by {Bell}{'}s theorem}},
	journal = {Nature},
	volume = {464},
	number = {7291},
	pages = {1021--1024},
	year = {2010},
	month = apr,
	issn = {1476-4687},
	publisher = {Nature Publishing Group},
	doi = {10.1038/nature09008}
}

@article{Reichardt2013,
	author = {Reichardt, Ben W. and Unger, Falk and Vazirani, Umesh},
	title = {{Classical command of quantum systems}},
	journal = {Nature},
	volume = {496},
	number = {7446},
	pages = {456--460},
	year = {2013},
	month = apr,
	issn = {1476-4687},
	publisher = {Nature Publishing Group},
	doi = {10.1038/nature12035}
}

@article{Tucker2024quantum,
  title={Quantum-assisted rendezvous on graphs: explicit algorithms and quantum computer simulations},
  author={Tucker, Joshua and Strange, Paul and Mironowicz, Piotr and Quintanilla, Jorge},
  journal={New Journal of Physics},
  volume={26},
  number={9},
  pages={093038},
  year={2024},
  publisher={IOP Publishing}
}

@article{Watts1998,
	author = {Watts, Duncan J. and Strogatz, Steven H.},
	title = {{Collective dynamics of {`}small-world{'} networks}},
	journal = {Nature},
	volume = {393},
	number = {6684},
	pages = {440--442},
	year = {1998},
	month = jun,
	issn = {1476-4687},
	publisher = {Nature Publishing Group},
	doi = {10.1038/30918}
}

@online{IBMQuantum_Kingston,
  author       = {{IBM Quantum}},
  title        = {{IBM} Quantum Processor: ibm\_kingston ({Heron} r2 Architecture)},
  howpublished = {\url{https://quantum.cloud.ibm.com/computers?processorType=Heron&system=ibm_kingston}},
  note={Accessed 12 June 2026}
}

%\newpage

\section{Methods}

\subsection{Discovering optimal strategies}

For a fixed graph, the number of actions per player equals the sum of all node degrees in the graph. For example, $C_3$ has three nodes of degree 2, providing 6 possible actions. 
For a graph ensemble, the number of actions per player equals the sum of all node degrees in the graph with the largest number of nodes (assuming equivalent nodes in different graphs have the same degrees). For nodes where local information allows players to determine the graph they are on, additional actions become available. 

A strategy consists of predetermined actions for every player and every reachable node. We build winning probability functions by mapping all potential routes from all possible starting locations.

\subsubsection{Classical Strategies}

Optimal strategies in our games are always deterministic as the probability of winning is a linear function of probabilities on all nodes, which are subject to linear constraints \cite{PhysRevA.109.042201}. This allows us to search through lists of binary variables that represent deterministic strategies. Let us explicitly go through the symmetric-rendezvous game on the ensemble \{$C_3$ , $C_6$\}. 

When the game is played on a single, known graph (e.g $C_3$), the list of binary digits has a length equal to the number of nodes. For instance, 
\begin{equation}
S(C_3) = (0,0,0)
\end{equation}
represents the ``go-to-lowest'' strategy. For graphs with larger numbers of nodes (e.g. $C_6$) the list grows accordingly. For instance,
\begin{equation}
    S(C_6) = (0,0,0,1,0,0)
\end{equation}
 represents the optimal classical strategy of go-to-lowest on all nodes except the fourth in which players go to the highest-indexed node available to them. In ensembles, the size of the list depends on both the number of nodes on the largest graph and the local information available to the players. If players cannot see the labels of neighbouring sites the size of the list is simply the number of nodes on the largest graph in the ensemble. For instance, 
\begin{equation}
    S(\{C_3%^{50 \%}
    ,%\cap 
    C_6%    ^{50 \%}
    \}) =  (0,0,0,1,0,0).
\end{equation}
represents a strategy where coinciding with $S(C_3)$ and $S(C_6)$ when players land on the respective graphs.  

As the amount of local information increases, the players have additional actions on sites that are shared by graphs in the ensemble. For instance, 
\begin{equation}
    S'(\{C_3%^{50 \%}
    ,%\cap 
    C_6%    ^{50 \%}
    \}) = (0,1,0,0,1,1,0,0)
\end{equation}
uses two additional Boolean variables to represent the distinct actions the players can take on the shared sites where they are able  to deduce the graph they are on (namely, sites 1 and 3).\\

The above discussion assumes player-symmetric strategies. A player-asymmetric strategy (where each player can follow a different set of instructions) can be encoded by concatenating $n$ lists like the ones above, where $n$ is the number of players. 

\subsubsection{Quantum strategies}
 Our optimal strategies start with the maximally entangled Bell state,
\begin{equation}
    |\psi\rangle_i = \frac{1}{\sqrt{2}}(|0\rangle_A|0\rangle_B + |1\rangle_A|1\rangle_B),
\end{equation}
where each player receives a qubit each. Because all Bell states are maximally entangled, they are interchangeable. Players need only adapt their strategy to the specific state selected. More general quantum states were also investigated but they offer no additional advantage (see Supplementary Information).

From their randomized starting nodes, players apply local operations to their respective qubits,
\begin{equation}
   \ket{\psi}_f =  R(\theta_a) \otimes R(\theta_b)\ket{\psi}_i = \frac{1}{\sqrt{2}} \begin{pmatrix}\cos(\frac{\theta_{b}-\theta_{a}}{2}) \\
        -\sin(\frac{\theta_{a}-\theta_{b}}{2}) \\
         \sin(\frac{\theta_{a}-\theta_{b}}{2}) \\
        \cos(\frac{\theta_{b}-\theta_{a}}{2})\\ 
    \end{pmatrix}
\end{equation}
where $R(\theta) = \exp(-i\frac12 \theta \hat{\sigma}_{y})$. %Alice and Bob apply their local rotations before measuring their respective qubits.

Since our figures of merit (such as rendezvous winning probability or graph domination number) are linear combinations of probabilities, it is straight-forward to construct general objective functions from the square modulus of the corresponding wave function components. These are then optimized using differential evolution \cite{storn1997differential} and L-BFGS with randomized starting parameters \cite{doi:10.1137/0916069}.
The strategy $\Sigma_q^1$ used for the simulations on IBM Kingston is given by the rotation angles $\theta_i = (5.50990981, 0, 1.25319582, 3.76853489, 0, 0)_i$, where $i$ is the site index.
The strategy $\Sigma_q^2$ used in the ion trap experiments is given by the rotation angles $\theta_i = (\alpha, \pi - \alpha, 0, 3\pi/4, -3\pi/4, \pi)_i$ with $\alpha \approx -1.93657837$.

%\subsection{General quantum states}

\subsection{NISQ simulations}

Simulations using NISQ hardware utilize the quantum table approach \cite{Tucker2024quantum,WeeksPREPRINT}: every possible action is mapped to an angle-parameterized quantum circuit. 20,000 shots of each circuit are run and the empirical outcomes are stored in a table. The classical simulation then proceeds by randomly sampling from these pre-computed tables.%  instead of repeatedly accessing the NISQ device.

\clearpage
\appendix

%\begin{center}
    \part{\begin{center}Supplementary Information\end{center}}
%\end{center}

\section{General quantum states}

We generalised our quantum strategies to more general shared quantum states:
\begin{equation}
    |\psi\rangle_i = \alpha|00\rangle + \beta|01\rangle  +\delta|10\rangle + \gamma|11\rangle
\end{equation}
 Here $\alpha,\beta,\delta,\gamma$ are complex numbers which we treat as additional optimization parameters, together with the angles of rotation around the Y axis (subject to normalisation). In all cases (for both Rendezvous and Graph Domination) the optimal quantum strategy had a maximally-entangled shared state.

\section{Rendezvous on graph ensembles with 9-site graphs}

\begin{figure}[!ht]
    \centering
        \includegraphics[width=0.98\linewidth]{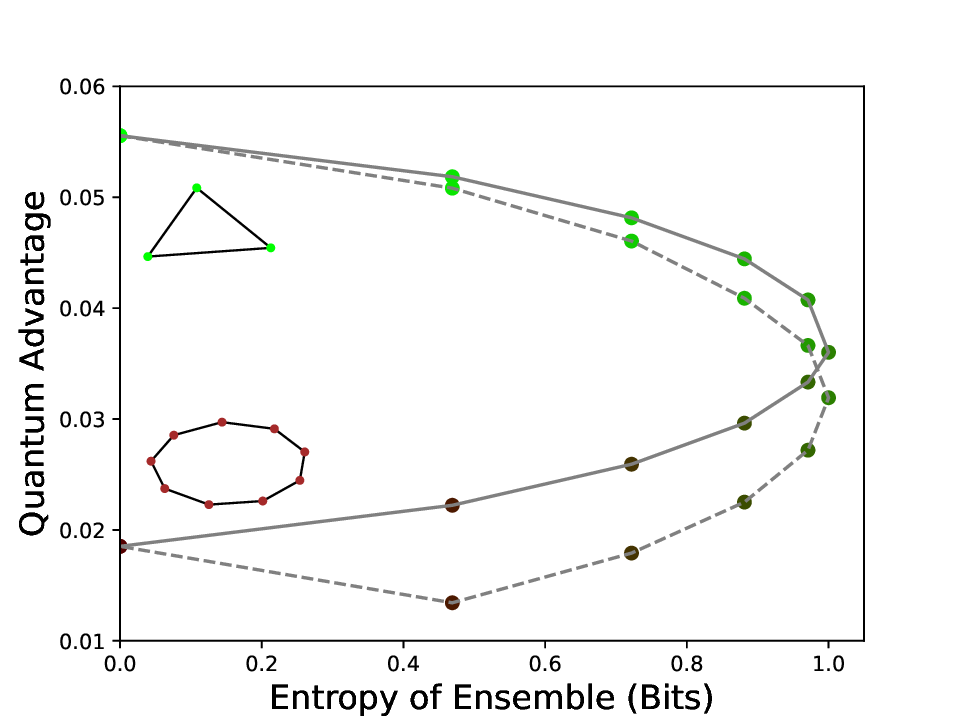}%
    \caption{Calculated quantum advantage $Q$ [Eq.~\eqref{eq:quantum_advantage_rendezvous}] for two-player, one-step rendezvous in an uncertain topography, represented by a statistical ensemble of the graphs  \{$C_3$,$C_9$\}. The conventions are the same is in Fig.~\ref{fig:EntropyAgainstAdvantageTwoGraphEnsembles} in the main text.}
    \label{fig:Rendezvous_C3C9}
\end{figure}

In the main text our studies of rendezvous on graph ensembles do not include graphs with more than seven sites. Fig.~\ref{fig:Rendezvous_C3C9} shows results for ensembles containing the graphs \{$C_3$,$C_9$\}.

\section{Successor/predecessor game variants}
In the main text we assumed that a player always knows which of the two edges they can take leads to the higher-index site, and which one leads to the lower-indexed site, of the two available.  An alternative would have been to let the player on site $i$ know which edge leads to site $i+1\mod N$ (successor) and which to site  $i-1\mod N$ (predecessor), where $N$ is the total number of nodes on the graph. 

When the ensemble contains only one graph, the two formulations are equivalent. In contrast, with topographic uncertainty they represent distinct versions of the game. The difference is in the amount of local information the players receive. 

To illustrate the above claim, let us consider a player starting on site 3 in an ensemble containing the graphs $C_3$ and $C_6$. In the successor/predecessor game, the player knows with certainty that if they take  the edge labelled  $i-1\mod N$ it will take them to site 2. In contrast, in the highest/lowest formulation they cannot predict with certainty the site index of the destination site for any of the two available edges. As a result, we find that the best strategies for successor/predecessor games always either match or beat the best strategies for highest/lowest games. This difference disappears if we allow players to see the labels of the nodes they are connected to.

\begin{figure}[H]
    \centering
    \includegraphics[width=0.5\textwidth]{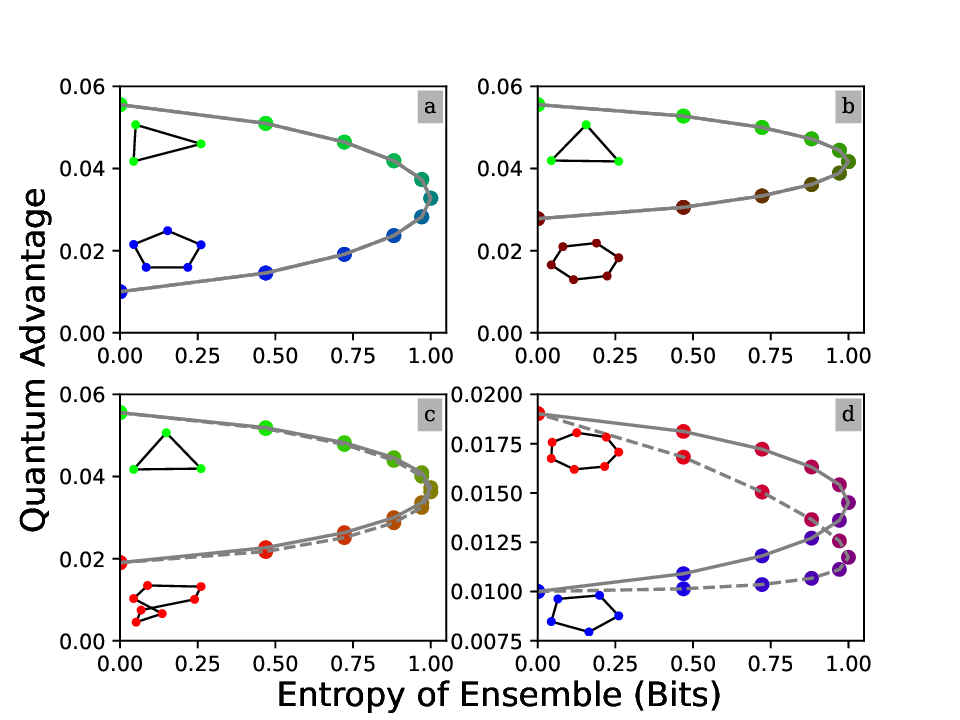}\\[1em] % <-- extra space
    \caption{Calculated quantum advantage for two-player, one-step rendezvous in an uncertain topography, represented by a statistical ensemble of the graphs  \{$C_3$,$C_5$\} (a), \{$C_3$,$C_6$\} (b), \{$C_3$,$C_7$\} (c), \{$C_5$,$C_7$\} (d) using the successor/predecessor formulation. It shows similar behaviour to the highest/lowest variant of the game but with less/no increase in quantum advantage when players are able to see neighbouring node labels. This is a consequence of players having intrinsically more local information in the successor/predecessor formulation of the game.\label{fig:rendezvousForwardBack} }
\end{figure}

\begin{figure}[H]
    %This figure needs to be updated to use the new definition of quantum advantage for this paper as I believe it currently does not. - DONE
    \centering
    \includegraphics[width=0.4\textwidth]{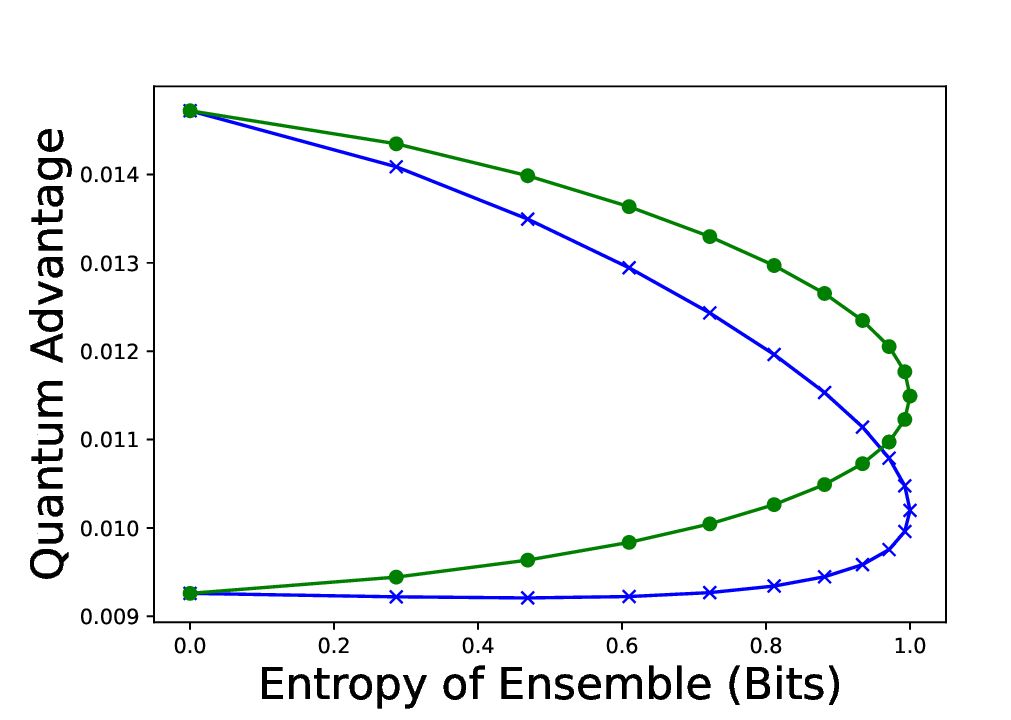}\\[1em] % <-- extra space
    \caption{Quantum advantage for a one step game of graph domination on ensembles of the graphs \{$C_5$,$C_6$\} using the successor/predecessor game formulation. The crosses correspond to the case with no additional site labels, while the filled circles correspond to the case where additional local information is available (signposts).}
    \label{fig:C5C6EnsembleMaxEntangledEntropyForwardBack}
\end{figure}

Figs.~\ref{fig:rendezvousForwardBack},\ref{fig:C5C6EnsembleMaxEntangledEntropyForwardBack} shows how the data in panels (a-d) of Fig.~\ref{fig:EntropyAgainstAdvantageTwoGraphEnsembles} in the main text change when the successor/predecessor variant of the rendezvous game is adopted. 


\section{Rendezvous C3 C6 Equations}
We start with the general definition of the average probability of winning a game of rendezvous on an ensemble containing $C_3$ and $C_6$,
\begin{align*}
     P_W = &P(C_3)P(W|C_3) + P(C_6)P(W|C_6)=\\ &P(C_3) \sum_{a,b=1}^3 \sum_{m,n=0}^1 P(W|a,b,m,n)\\
     &+ P(C_6)\sum_{c,d=1}^6 \sum_{m,n=0}^1 P(W|c,d,m,n).
\end{align*}
Here $P(W|C_3)$ and $P(W|C_6)$ represent the probabilities of winning a game of rendezvous using a fixed joint strategy, given that the players are placed on $C_3$ and $C_6$, respectively, and $P(C_n)$ is the ensemble probability of graph $C_n$. When players are unable to see the labels of the neighbouring nodes they are connected to, a strategy is specified by the angles $\{ \theta_{A,i} , \theta_{B,i} \}_{i=1}^6$ that Alice and Bob will rotate their qubits by at each site. The conditional probabilities are then given explicitly by
\begin{align*}
P(W|C_3) = \frac{1}{9} \bigg[ 
    &\sum_{i=1}^3 \cos^2 \left( \tfrac{\theta_{Ai} - \theta_{Bi}}{2} \right) \\
    + &\cos^2\left(\tfrac{\theta_{A1} - \theta_{B2}}{2}\right) + \cos^2\left(\tfrac{\theta_{A2} - \theta_{B1}}{2}\right) \\
    + &\sin^2\left(\tfrac{\theta_{A1} - \theta_{B3}}{2}\right) + \sin^2\left(\tfrac{\theta_{A3} - \theta_{B1}}{2}\right) \\
    + &\cos^2\left(\tfrac{\theta_{A2} - \theta_{B3}}{2}\right) + \cos^2\left(\tfrac{\theta_{A3} - \theta_{B2}}{2}\right) \bigg]
\end{align*}%
and %
\begin{align*}
P(W|C_6) = \frac{1}{36} \bigg( 
    &\sum_{i=1}^6 \cos^2 \left( \tfrac{\theta_{Ai} - \theta_{Bi}}{2} \right) \\
    + \frac{1}{2} \Big[ 
    &\cos^2\left(\tfrac{\theta_{A1} - \theta_{B2}}{2}\right) + \cos^2\left(\tfrac{\theta_{A2} - \theta_{B1}}{2}\right) \\
    + &\cos^2\left(\tfrac{\theta_{A1} - \theta_{B3}}{2}\right) + \cos^2\left(\tfrac{\theta_{A3} - \theta_{B1}}{2}\right) \\
    + &\cos^2\left(\tfrac{\theta_{A1} - \theta_{B5}}{2}\right) + \cos^2\left(\tfrac{\theta_{A5} - \theta_{B1}}{2}\right) \\
    + &\sin^2\left(\tfrac{\theta_{A1} - \theta_{B6}}{2}\right) + \sin^2\left(\tfrac{\theta_{A6} - \theta_{B1}}{2}\right) \\
    + &\sin^2\left(\tfrac{\theta_{A2} - \theta_{B3}}{2}\right) + \sin^2\left(\tfrac{\theta_{A3} - \theta_{B2}}{2}\right) \\
    + &\sin^2\left(\tfrac{\theta_{A2} - \theta_{B4}}{2}\right) + \sin^2\left(\tfrac{\theta_{A4} - \theta_{B2}}{2}\right) \\
    + &\cos^2\left(\tfrac{\theta_{A2} - \theta_{B6}}{2}\right) + \cos^2\left(\tfrac{\theta_{A6} - \theta_{B2}}{2}\right) \\
    + &\sin^2\left(\tfrac{\theta_{A3} - \theta_{B4}}{2}\right) + \sin^2\left(\tfrac{\theta_{A4} - \theta_{B3}}{2}\right) \\
    + &\sin^2\left(\tfrac{\theta_{A3} - \theta_{B5}}{2}\right) + \sin^2\left(\tfrac{\theta_{A5} - \theta_{B3}}{2}\right) \\
    + &\sin^2\left(\tfrac{\theta_{A5} - \theta_{B4}}{2}\right) + \sin^2\left(\tfrac{\theta_{A4} - \theta_{B5}}{2}\right) \\
    + &\cos^2\left(\tfrac{\theta_{A4} - \theta_{B6}}{2}\right) + \cos^2\left(\tfrac{\theta_{A6} - \theta_{B4}}{2}\right) \\
    + &\cos^2\left(\tfrac{\theta_{A6} - \theta_{B5}}{2}\right) + \cos^2\left(\tfrac{\theta_{A5} - \theta_{B6}}{2}\right) \Big] \bigg),
\end{align*}
respectively. Further simplifcation can be achieved by noting that the optima strategy is player symmetric, i.e. $\theta_{A,i}=\theta_{B,i}$ for all $i.$ This gives
\begin{align*}
    P(W|C_3) = \frac{1}{9} \Big( 3 + 2 \big[ 
    &\cos^2\left(\tfrac{\theta_2 - \theta_1}{2}\right) \\
    + &\cos^2\left(\tfrac{\theta_3 - \theta_2}{2}\right) \\
    + &\sin^2\left(\tfrac{\theta_3 - \theta_1}{2}\right) \big] \Big)
\end{align*}
and
\begin{align*}
    P(W|C_6) = \frac{1}{36} \Big( 6 
    &+ \cos^2\left(\tfrac{\theta_2 - \theta_1}{2}\right) + \sin^2\left(\tfrac{\theta_3 - \theta_2}{2}\right) \\
    &+ \sin^2\left(\tfrac{\theta_4 - \theta_3}{2}\right) + \sin^2\left(\tfrac{\theta_5 - \theta_4}{2}\right) \\
    &+ \sin^2\left(\tfrac{\theta_6 - \theta_1}{2}\right) + \cos^2\left(\tfrac{\theta_5 - \theta_6}{2}\right) \\
    &+ \cos^2\left(\tfrac{\theta_3 - \theta_1}{2}\right) + \sin^2\left(\tfrac{\theta_4 - \theta_2}{2}\right) \\
    &+ \sin^2\left(\tfrac{\theta_3 - \theta_5}{2}\right) + \cos^2\left(\tfrac{\theta_6 - \theta_2}{2}\right) \\
    &+ \cos^2\left(\tfrac{\theta_6 - \theta_4}{2}\right) + \cos^2\left(\tfrac{\theta_5 - \theta_1}{2}\right) \Big).
\end{align*}
For the particular case of a uniformly distributed ensemble  $\left[P(C_3) = P(C_6) = \frac{1}{2}\right]$ this yields 
\begin{align*}
P_W = \frac{5}{18} + \frac{1}{72} \bigg[ 
    & 9 \cos^2\left(\tfrac{\theta_2 - \theta_1}{2}\right) + 7 \cos^2\left(\tfrac{\theta_3 - \theta_2}{2}\right) \\
    + & 7 \sin^2\left(\tfrac{\theta_3 - \theta_1}{2}\right) + \sin^2\left(\tfrac{\theta_4 - \theta_3}{2}\right) \\
    + & \sin^2\left(\tfrac{\theta_5 - \theta_4}{2}\right) + \sin^2\left(\tfrac{\theta_6 - \theta_1}{2}\right) \\
    + & \cos^2\left(\tfrac{\theta_5 - \theta_6}{2}\right) + \sin^2\left(\tfrac{\theta_4 - \theta_2}{2}\right) \\
    + & \sin^2\left(\tfrac{\theta_3 - \theta_5}{2}\right) + \cos^2\left(\tfrac{\theta_6 - \theta_2}{2}\right) \\
    + & \cos^2\left(\tfrac{\theta_6 - \theta_4}{2}\right) + \cos^2\left(\tfrac{\theta_5 - \theta_1}{2}\right) \bigg].
\end{align*}

One can similarly derive equations for the case when players know the site labels of nearest neighbours. The main difference is that each player has available two distinct angles for sites 1 and two more for site 3, depending on the inference they have made regarding the graph on which they game is being played:
\[
\begin{aligned}
\theta_{A1},\theta_{A3},\theta_{B1},\theta_{B3}
&\to \theta_{A1}^{C_3},\theta_{A1}^{C_6}, \theta_{A3}^{C_3},\theta_{A3}^{C_6}, \\
&\quad\quad \theta_{B1}^{C_3},\theta_{B1}^{C_6}, \theta_{B3}^{C_3},\theta_{B3}^{C_6}.
\end{aligned}
\]
On other sites, the players either remain ignorant of the graph (2), or are certain even without the additional labels (4,5,6).

\section{Domination C5 C6 Equations}
The average domination number of the an ensemble containing the graphs $C_5$,$C_6$ is formed by two contributions, one for each graph in the ensemble:
\begin{equation}
    D=P\left(C_5\right)D_5+P\left(C_6\right)D_6.
    \label{eq:D_two_graphs}
\end{equation}
The individual domination numbers $D_n$ ($n=5,6$) are given by 
%
%\begin{widetext}
\begin{align}
D_n={}&
\sum_{i,j=1}^{n}
\left[
A^n_{ij}
\sin^{2}\!\left(\frac{\theta_{A_i}-\theta_{B_j}}{2}\right)
\right.
\nonumber\\
&\hspace{2.4cm}\left.
+
B^n_{ij}
\cos^{2}\!\left(\frac{\theta_{A_i}-\theta_{B_j}}{2}\right)
\right],
\label{eq:D_one_graph}
\end{align}
%\end{widetext}
where the coefficient matrices $A^n,B^n$ are given by 
\begin{widetext}
\begin{equation}
A^5 \equiv 
\left(
\begin{array}{ccccc}
5 & 4 & 5 & 5 & \frac{9}{2}\\
4 & 5 & \frac{9}{2} & \frac{7}{2} & 5\\
5 & \frac{9}{2} & 5 & \frac{9}{2} & 5\\
5 & \frac{7}{2} & \frac{9}{2} & 5 & 4\\
\frac{9}{2} & 5 & 5 & 4 & 5
\end{array}
\right);~
B^5 \equiv 
\left(
\begin{array}{ccccc}
3 & \frac{9}{2} & \frac{7}{2} & \frac{7}{2} & 4\\
\frac{9}{2} & 3 & 4 & 5 & \frac{7}{2}\\
\frac{7}{2} & 4 & 3 & 4 & \frac{7}{2}\\
\frac{7}{2} & 5 & 4 & 3 & \frac{9}{2}\\
4 & \frac{7}{2} & \frac{7}{2} & \frac{9}{2} & 3
\end{array}
\right);~
A^6={}
\left(
\begin{array}{cccccc}
5 & 4 & 5 & 6 & 5 & 5\\
4 & 5 & 5 & 4 & 4 & 5\\
5 & 5 & 5 & 5 & 4 & 6\\
6 & 4 & 5 & 5 & 5 & 5\\
5 & 4 & 4 & 5 & 5 & 4\\
5 & 5 & 6 & 5 & 4 & 5
\end{array}
\right);~
B^6=\left(
\begin{array}{cccccc}
3 & 5 & 4 & 4 & 4 & 4\\
5 & 3 & 4 & 5 & 6 & 4\\
4 & 4 & 3 & 4 & 5 & 4\\
4 & 5 & 4 & 3 & 4 & 4\\
4 & 6 & 5 & 4 & 3 & 5\\
4 & 4 & 4 & 4 & 5 & 3
\end{array}
\right).
\label{eq:D_numbers}
\end{equation}
\end{widetext}

The coefficients in (\ref{eq:D_numbers}) were determined by the following procedure: i) place players at the specified locations; ii) make the corresponding moves ($\cos^2$ terms correspond to players making the same moves from their respective sites, while $\sin^2$ terms correspond to players taking opposite actions); iii) determine the resulting domination number. We then use (\ref{eq:D_one_graph}) to compute average domination for each graph and (\ref{eq:D_two_graphs}) for the average domination across the ensemble. Everything else is deduced in the same way as for the rendezvous problem, and players' knowledge of their surroundings is implemented identically as well.

%We are able to look at how the angles for a given strategy change as we vary the probability of them being placed onto either of the graphs

\section{Increase in rendezvous probability with more local information }

\begin{figure}[H]
    \centering
    \includegraphics[width=0.5\textwidth]{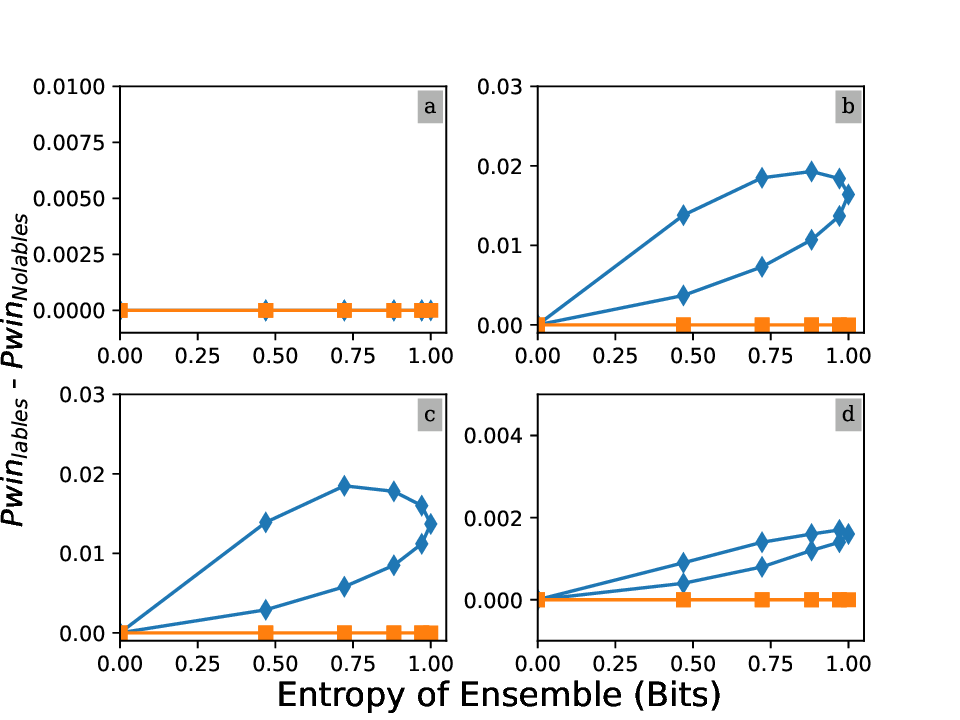}
    \caption{Change in the optimal probability for rendezvous when additional labels are provided for classical (squares) and quantum (diamonds) strategies. Each plot corresponds to an ensemble involving a different set of 
    graphs: \{$C_3$,$C_5$\} (a), \{$C_3$,$C_6$\} (b), \{$C_3$,$C_7$\} (c), and \{$C_5$,$C_7$\} (d). The horizontal axis shows the entropy of a given ensemble.}
    \label{fig:probability_comparison_information}
\end{figure}

Fig.~\ref{fig:probability_comparison_information} shows the increase in rendezvous probability when more local information is available for optimal classical and quantum strategies. 

Panels (b-d), corresponding to ensembles involving \{$C_3$,$C_6$\}, \{$C_3$,$C_7$\}, and \{$C_5$,$C_7$\}, show an increase for all finite-entropy ensembles in the case of quantum strategies. In contrast, classical strategies do not gain from the additional information. In panel (a), corresponding to \{$C_3$,$C_5$\}, neither classical not quantum strategies show an improvement. This further illustrates that quantum strategies are able to make use of additional local information to increase the probability of winning, the only exception being when a strategy that is already optimal strategy for all graphs in the ensemble can be found [the case of panel (a)].

\section{How Angles Change over the distribution plot }

\begin{figure}[H]
    \centering
    \includegraphics[width=0.5\textwidth]{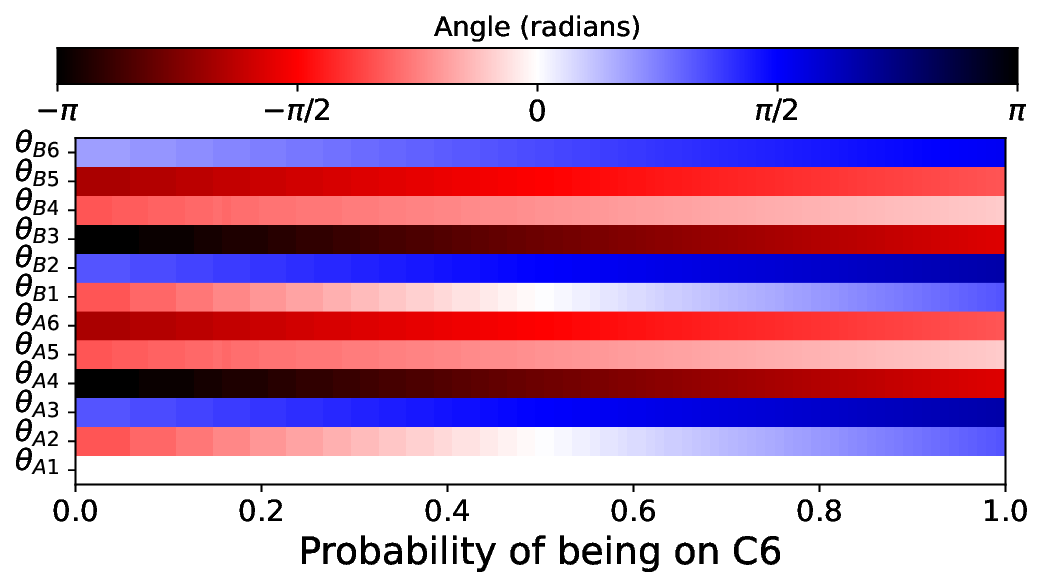}
    \caption{Optimal quantum strategy for graph ensembles containing the cycles $C_3$ and $C_6.$ The particular ensemble chosen varies along the horizontal axis and is identified by the probability of the larger graph in the ensemble. 101 ensembles were investigated, obtained by increasing the probability in steps of 0.01. The vertical axis labels the angles characterising the strategy (for their definitions, see the main text). The colour map shows the value of each angle for each particular ensemble.}
    \label{fig:AngleDistirbution}
\end{figure}

Fig.~\ref{fig:AngleDistirbution} shows how optimal quantum rendezvous strategies evolve as we vary the composition of the graph ensemble in which the game is played. 

%\section{Domination Graph}

\end{document}